\documentclass[%
aps,
prl,
reprint,
superscriptaddress,
showpacs,preprintnumbers,
amsmath,amssymb
]{revtex4-2}

\pdfoutput=1 
\usepackage{graphicx}   
\usepackage{dcolumn}
\usepackage{bm}
\usepackage{hyperref}
\usepackage{braket}
\usepackage{siunitx}
\usepackage[utf8]{inputenc}
\usepackage{xcolor} 

\begin{document}
	
\title{Random telegraph fluctuations in granular microwave resonators}

\author{M. Kristen}	
\affiliation{Institute for Quantum Materials and Technology, Karlsruhe Institute of Technology, 76344 Eggenstein-Leopoldshafen, Germany}	
\affiliation{Institute for Quantum Materials and Technologies, 76131 Karlsruhe, Germany}
\author{J. N. Voss}	
\author{M. Wildermuth}		
\affiliation{Institute for Quantum Materials and Technologies, 76131 Karlsruhe, Germany}
\author{H. Rotzinger}
\email{rotzinger@kit.edu}	
\author{A. V. Ustinov}	
\affiliation{Institute for Quantum Materials and Technologies, Karlsruhe Institute of Technology, 76344 Eggenstein-Leopoldshafen, Germany}	
\affiliation{Institue of Physics, Karlsruhe Institute of Technology, 76131 Karlsruhe, Germany}

\date{\today}

\begin{abstract}
Microwave circuit electrodynamics of disordered superconductors is a very active research topic spawning a wide range of experiments and applications. For compact superconducting circuit elements, the transition to an insulating state poses a limit to the maximum attainable kinetic inductance. It is therefore vital to study the fundamental noise properties of thin films close to this transition, particularly in situations where a good coherence and temporal stability is required. In this paper, we present measurements on superconducting granular aluminum microwave resonators with high normal state resistances, where the influence of the superconductor to insulator phase transition is visible. We trace fluctuations of the fundamental resonance frequency and observe, in addition to a $1/f$ noise pattern, a distinct excess noise, reminiscent of a random telegraph signal. The excess noise shows a strong dependency on the resistivity of the films as well as the sample temperature, but not on the applied microwave power.

\end{abstract}


\maketitle

The  phase transition from a superconducting to an insulating state (SIT) of disordered thin films remains under intense debate \cite{Dubi2007,Sacepe2020}. The prevailing interest in the various aspects of this transition is owed to the intrinsic disorder of high-Tc superconductors \cite{Bollinger2011,Harris2018,Zhou2022}, as well as the use of disordered superconductors in quantum circuits and particle detectors \cite{Doucot2012,Baselmans2012,Zmuidzinas2012}. 

Generally, the breakdown of the superconducting state manifests itself in the suppression of the long-range order parameter $\varPsi = \Delta \mathrm{e}^{\mathrm{i}\phi}$. In the case of granular systems \cite{Beloborodov2007} it is believed that, while the amplitude $\Delta$ persists, the stiffness \cite{Pracht2016} of the phase $\phi$ is lost when the effective Coulomb energy surpasses the energy of the Josephson coupling $E_\mathrm{J} \propto \Delta / R_\mathrm{n} $ between neighboring grains \cite{ Anderson1964, Abeles1977, Efetov1980}. This means that Cooper pairs, the charge carriers of the superconducting state, can no longer tunnel coherently between grains and the superconducting behavior of the whole sample is suppressed \cite{Humbert2021, Voss2021}. In agreement with theoretical predictions \cite{Chakravarty1987}, experiments have shown that this coincides with a normal state sheet resistance $R_\mathrm{n}$ on the order of the superconducting resistance quantum $R_q = 6.45\,\rm{k\Omega}$ \cite{Jaeger1989}. 

Disordered films in the vicinity of the SIT show a variety of intriguing physical effects, like charge localization or subgap absorption. Experimental means to study such phenomena include scanning tunneling microscopy \cite{Bouadim2011,Sacepe2011,Sherman2015,Dubouchet2019,Yang2020}, optical spectroscopy \cite{Pracht2012,Sherman2014, Pracht2016, Bertrand2019} or transport measurements \cite{Dynes1984,Jaeger1986,Han2014, Roy2020}.  While these techniques offer unique insights into the rich SIT physics, they provide only limited information regarding the applicability of these materials in high impedance microwave circuits, where they are sought after due to their sizable kinetic inductance $L_\textrm{k} \propto R_\mathrm{n}$. 

In this work, we attempt to bridge this knowledge gap through a detailed study on the low frequency noise properties of compact, highly resistive superconducting granular aluminum microwave resonators \cite{Rotzinger2017}. We observe pronounced fluctuations of the resonance frequency at temperatures of $10-200\,\mathrm{mK}$, which intensify in samples with a higher normal-state resistance. In contrast to conventional aluminum resonators, the $1/f$ noise spectrum of granular aluminum is substantially higher and masked by random telegraphic signal (RTS) like fluctuations. While the amplitude $I$ of the RTS is independent of the measurement power and temperature, the RTS switching time $\tau_0$ abruptly decreases above $200 \, \textrm{mK}$.

\begin{table}[b]
	\caption{\label{tab1}Characteristics of the measured samples. All resonators have a width of $2 \, \si{\micro\metre}$, but vary in their lenght $l$. The average resonance frequency $f_0$ and resonator linewidth $\kappa$ are extracted from a fit to the resonance \cite{Probst2015}. $R_\mathrm{n}$ is the normal conducting sheet resistance of the films.}
	\begin{ruledtabular}
		\begin{tabular}{cccccc}
			Resonator & $f_0 \, ({\rm GHz})$ & $R_\mathrm{n} \, ({\rm k\Omega/\square})$ & $l \, \textrm{(\si{\micro\metre})}$ & $ \kappa \, ({\rm MHz})$  \\ 
			A1 & 10.565 & 0.6 & 406 & 2.33 \\
			B1 & 5.494 & 1.4 & 505 &  1.84\\
			B2 & 6.154 & 1.5 & 440 & 2.23 \\
			B3 & 6.793 & 1.5 & 390 & 2.61 \\
			C1 & 4.069 & 4.0 & 406 & 0.61 \\
			C2 & 4.663 & 4.3 & 337 & 0.82 \\
			C3 & 5.780 & 3.8 & 287 & 1.51 \\	
		\end{tabular}
	\end{ruledtabular}
\end{table}

\begin{figure}
	\includegraphics{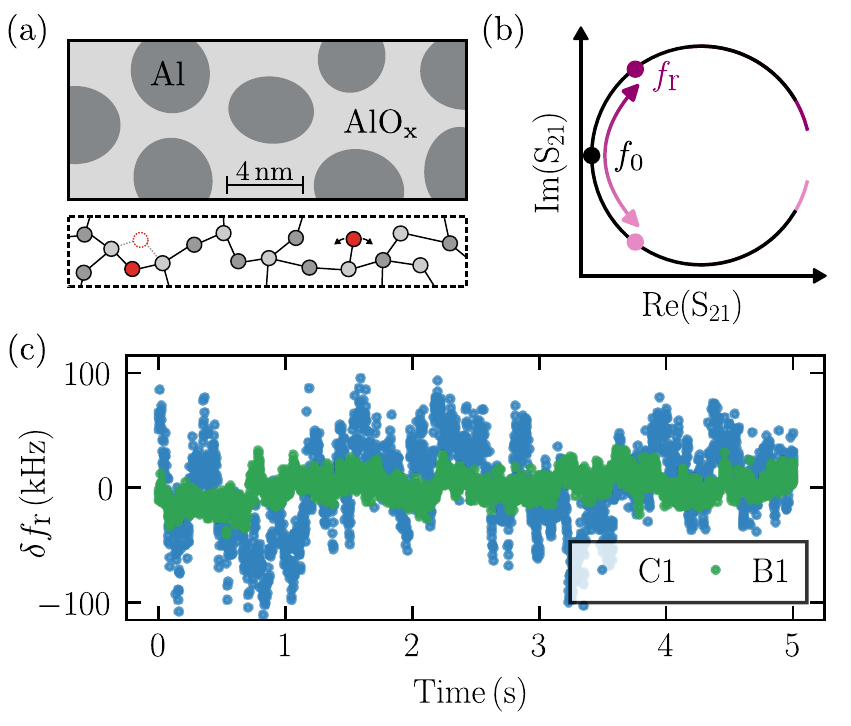}
	\caption{\label{Fig1} (a) Top panel: Schematic microstructure of granular aluminum. The grain dimension, location and the inter-grain AlO$_\textrm{x}$ barrier are subject to disorder \cite{Voss2021,Bartolo2022}. Bottom panel: Two-level defects (red) within the amorphous atomic structure of AlO$_\textrm{x}$. (b) Sketch of an ideal resonance circle in the complex plane of the transmission signal. A fluctuation of the resonator frequency $\delta f_\textrm{r} = f_\textrm{r} - f_\textrm{0}$ corresponds to a rotation of the circle. This change is monitored by continuously measuring $S_\textrm{21}$ with the probe frequency fixed to $f_0$. (c) Raw frequency fluctuations of resonator C1 and B1, recorded at an average photon number $\overline{n} \sim 10^4$.}    
\end{figure}

\begin{figure}
	\includegraphics{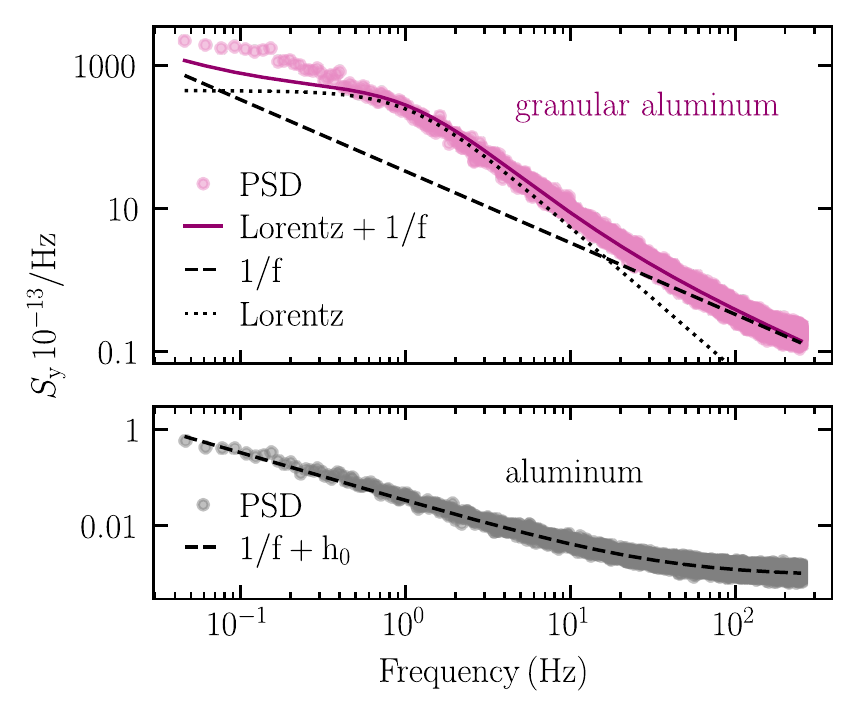}
	\caption{\label{Fig2} Low temperature ($T=10\, \rm mK$) fractional noise spectra of granular aluminum resonators (here: C1) show a $1/f$ dependency (dashed line), masked by RTS excess noise below 10\,Hz (dotted line). Solid line is a fit to Eq.~\ref{Eq1}. Pure aluminum films measured identically show no sign of RTS noise and orders of magnitude lower $1/f$ noise levels.}
\end{figure}

\begin{figure*}
	\includegraphics[width=\textwidth]{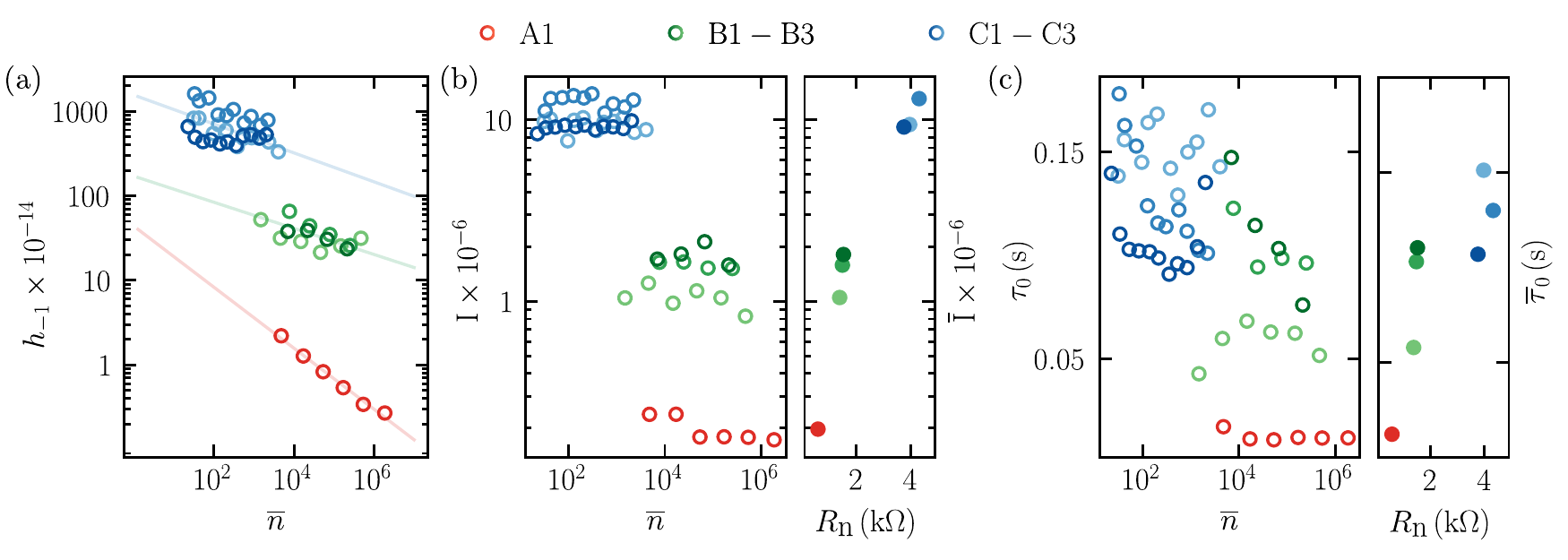}
	\caption{\label{Fig3} Power ($\overline{n} \propto P_\textrm{MW}$) dependence of the noise parameter. (a) The amplitude $h_{-1}$ of the 1/f noise shows a strong dependency on the number of photons in the resonator $\overline{n}$. Solids lines are a fit to $1/\overline{n}^\beta$. The amplitude (b)  and lifetime (c) of the RTS fluctuations are not affected by $\overline{n}$, but shows a noticeable dependency on the film resistance $R_\textrm{n}$ (right panel, respectively).}
\end{figure*}

\begin{figure}
	\includegraphics{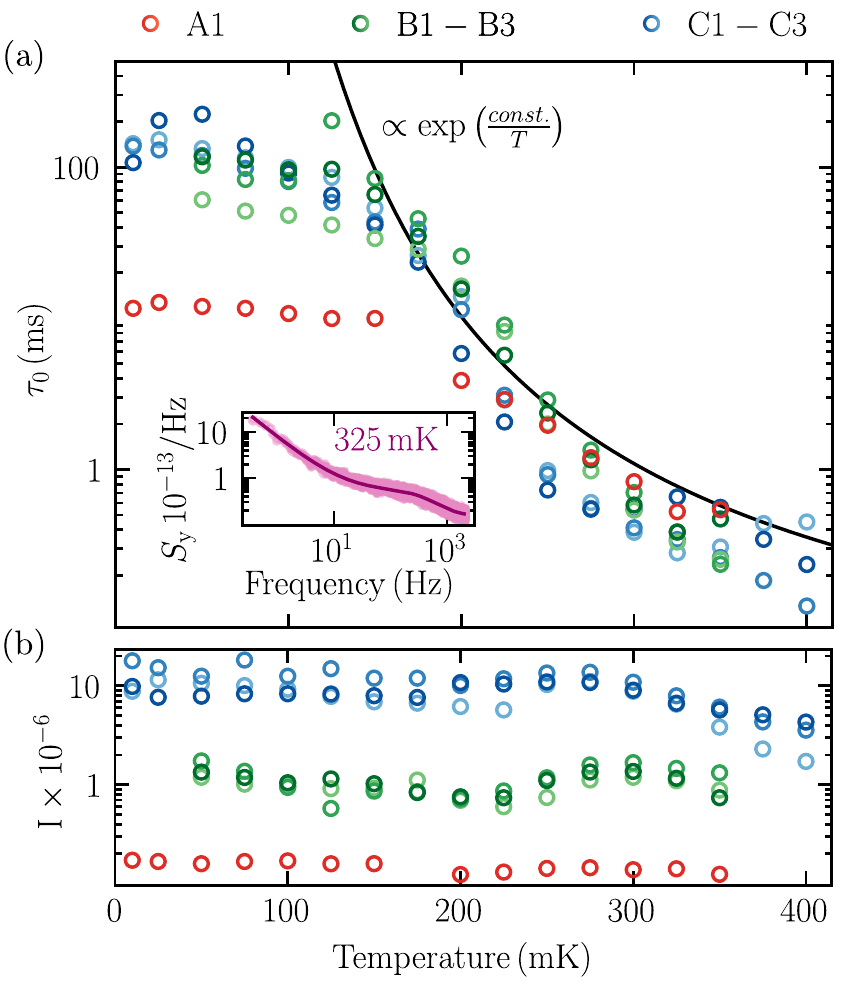}
	\caption{\label{Fig4} Temperature dependence of the RTS parameters. While the RTS lifetime $\tau_0$ decreases exponentially (solid line, not a fit) above $~200 \, \rm mK$ (a), the amplitude $A$ is almost independent of the temperature (b). Inserts shows the power spectral density of resonator C1 measured at $325 \, \rm mK$.}
\end{figure}


The microwave resonators (A1-C3, see Tab.~\ref{tab1}) have been fabricated from three $22-30 \, \textrm{nm}$ thick granular aluminum films with different sheet resistances (see Fig.~\ref{Fig1}(a) for schematic of its microstructure). The films were prepared on sapphire substrates by reactive sputter deposition of aluminum in an oxygen atmosphere, using an in-situ control of the sheet resistance \cite{Wildermuth2022}.  On each chip, multiple half wavelength microstrip resonators coupled to a common transmission line were structured using an optical resist mask and an anisotropic dry etching process. The chips were installed in microwave tight sample box and mounted to the mixing chamber plate of a dry dilution refrigerator, with experimental temperatures ranging from 10 to 400\,mK.  

The complex transmission coefficient $S_{21}$ in the vicinity of the resonators was measured using a vector network analyzer (VNA). To record the time-dependent frequency fluctuations, the probe frequency was fixed to the average resonator frequency $f_\textrm{0} \equiv \overline{f}_\textrm{r}$. If the resonator frequency changes by $\delta f_\textrm{r} = f_\textrm{r} - f_\textrm{0}$, the position of $S_{21}(f_0)$ in the complex plane proportionately shifts along the resonance circle (see Fig.~\ref{Fig1}(b)). Using knowledge of the pre-measured resonance circle, each newly measured value can then be mapped to a corresponding frequency $f_\textrm{r}$ (see supplementary materials for details). A full data sets contains $\mathcal{O}(10^6)$ of such measurements, taken at a rate $\geq 500\textrm{/s}$.  

Figure~\ref{Fig1}(c) shows extracts from mapped data obtained in typical noise measurements. Compared to the other resonators, the fluctuations of the resonance frequency $\delta f_\textrm{r}$ are much more pronounced in the resonators with the highest sheet resistances (C1-C3), measuring values up to $\delta f_\textrm{r}/\kappa=0.1$. Here, $\kappa$ is the individual resonator linewidth.

Our analysis of the frequency fluctuations focuses on the fractional noise spectra defined as $S_\textrm{y}= S_{\delta f_\textrm{r}}/f_\textrm{0}^2$ \cite{Burnett2014}, where the power spectral density $S_{\delta f_\textrm{r}}$ (in units of $\mathrm{Hz}^2/\mathrm{Hz}$) is calculated from the datasets using Welch's method \cite{Welch1967}. Fig.~\ref{Fig2} shows the noise spectrum of resonator C1 compared to the spectrum of a pure aluminum resonator ($R_\mathrm{n} \sim \SI{0.3}{\Omega}$) measured under identical conditions at $T = 10 \, \rm mK$.  For frequencies above 10\,Hz, both spectra follow a $1/f$ trend. All granular aluminum resonators show, however, orders of magnitude higher noise amplitudes.  Additionally, in the region between  $0.1 \, \rm Hz$ and $10 \, \rm Hz$, the spectrum noticeably deviates from the $1/f$ trend. The spectral shape of these low frequency excess fluctuations indicates an RTS, i.e., the resonator switches between frequency-distinct states.

Similar to previous works \cite{deVisser2011,Schlor2019, Burnett2019, Niepce2021}, we model their contribution to the spectrum by a Lorentzian centered at zero frequency. This corresponds to a randomly excited process that exponentially decays on a characteristic time scale $\tau_0$. Including a white noise floor, the full fractional noise spectrum can then be described by the expression

\begin{align}
	S_\textrm{y}= \frac{4I^2\tau_0}{1+(2\pi f \tau_0)^2}+\frac{h_{-1}}{f} + h_0, 
	\label{Eq1}
\end{align}

with the amplitude of the RTS $I$, the $1/f$ noise $h_{-1}$ and the white noise $h_0$, respectively. The following discussion is based on a least square fit of Eq. \ref{Eq1} to all measured noise spectra.

Figure~\ref{Fig3} shows the dependence of the fitting parameters on the average number of photons  $\overline{n}=2P_{\rm VNA}\kappa_{\rm c}/(\hbar \omega_{\rm r}^3\kappa^2)$ oscillating in the resonator \cite{Schneider2020}, which is controlled by the applied VNA power $P_{\rm VNA}$. 
The amplitude of the $1/f$ noise $h_{-1}$ shows a power law dependency, where a comparison to $h_{-1} \propto 1/\overline{n}^\beta$ yields $\beta = 0.36,0.15,0.17$ (Fig.~\ref{Fig3}(a)). However, no clear dependency on $\overline{n}$ can be observed for the parameters of the RTS, despite photon numbers spanning over several orders of magnitude. Note that applied VNA power above $\overline{n}\gg10^{4}$ leads to strong non-linear resonance bifurcations \cite{Swenson2013,He2021} in resonators C1-C3 and is therefore not taken into account. The missing data points of resonators A1 and B1-B3 at low photon numbers are due to an obscured (small) RTS signal at an increased $1/f$ amplitude ($h_{-1}/\SI{1}{\hertz} \gtrsim 4I^2\tau_0$). However, for $\overline{n} \sim 10^4$ the data points overlap and one can therefore compare the average values of the RTS amplitude $\overline{I}$ and lifetime $\overline{\tau_0}$ between resonators (Fig.~\ref{Fig3}(b)+(c), right panel). The comparison indicates a dependence on $R_\textrm{n}$, which agrees with the initial observations (Fig.~\ref{Fig1}(c)) that $\delta f_\mathrm{r}$ is most pronounced in resonators made from the most restive film C. 

Following the power scans, the dependence of the RTS characteristics on the sample temperature was investigated in the range from $10-400$\,mK. As shown in Fig.~\ref{Fig4}(a), the RTS lifetime $\tau_0$ decreases rapidly above a threshold temperature of $\sim 200\,\mathrm{mK}$ in all measured resonators. This drop is approximately exponential, as indicated by the black line. The fluctuation amplitude $I$, however, remains approximately constant over the whole temperature range (Fig.~\ref{Fig4}(b)). Note that $I \propto \delta\textrm{RTS} \times \delta f_\textrm{r}/\delta\textrm{RTS}$, with $\delta\textrm{RTS}$ the amplitude of the RTS process and $\delta f_\textrm{r}/\delta\textrm{RTS}$ proportional to the coupling between the RTS fluctuators and the resonator \cite{deVisser2012}. Both quantities may have an opposite temperature dependency which cancel out for the overall contribution to $I$.

The observed $1/f$ scaling of the frequency noise is a well-known phenomenon in thin-films that has been studied in widely different systems, revealing a variety of physical sources \cite{Kogan1996}. In superconducting microwave resonators, it is proposed to originate from electric dipole coupling to atomic defects behaving as two-level systems (TLS), e.g., accumulating in the surface oxide \cite{Paladino2014,Muller2019} (see Fig.~\ref{Fig1}(a)). 

Within the generalized tunneling model, these TLS are believed to interact with surrounding defects having interlevel transition frequencies below $k_\mathrm{B} T$ and are therefore subjected to thermal fluctuations \cite{Faoro2012, Burnett2014}. The model predicts that the corresponding $1/f$ noise amplitude $h_{-1}$ scales with $\beta = 0.5$ at high photon numbers and $\beta \rightarrow 0$ as the photon number decreases \cite{Faoro2015}, which qualitatively agrees with our findings. 
Further, $h_{-1}$ clearly depends on the sheet resistance of the resonator film, which is expected as the number of TLS increases with the thickness of the inter-grain oxide barrier. This complements the observed prevalence of strongly coupled TLS, which we discuss in a separate publication \cite{Kristen}.

In the light of this interpretations, it seems natural to also attribute the RTS noise component to TLS. Indeed, it has been shown in 'transmon' type superconducting qubits \cite{Schlor2019, Burnett2019} as well as superconducting resonators \cite{Niepce2021} that a nearly resonant TLS can produce a dominant Lorentzian noise spectrum. However, for TLS processes a reduction of the fluctuation amplitude $I$ with $\overline{n}$ similar to $h_{-1}(\overline{n})$ would be expected for the resonator-TLS system, which we do not observe \cite{Niepce2021}.  Further, due to the random nature of these defects, it is statistically unlikely to find TLS properties only varying between films, but not resonators. 
	
Noise measurements in narrow aluminum resonators showed that the creation and recombination of quasiparticles can also lead to RTS fluctuations \cite{deVisser2011,deVisser2012}. The measured $\tau_0$ values are comparable to quasiparticle lifetimes previously observed in granular aluminum \cite{Grunhaupt2018}. For quasiparticles, the exponential decrease of the lifetime depicted in Fig.~\ref{Fig4}(a) would be expected naturally, as their number $N_\textrm{qp}$ increases with temperature and it becomes more likely to find a pairing partner. Since the responsivity of the resonator to the quasiparticles $\delta f_\textrm{r}/\delta\textrm{RTS}$ is almost temperature independent \cite{Baselmans2008, Gao2008a} and $\delta\textrm{RTS} \propto N_\textrm{qp}$, the noise amplitude $I$ should instead increase with temperature. The data presented in Fig.~\ref{Fig4}(b) contradicts this assumption, where $I$ rather decreases with increasing temperature. In addition, we do not observe a broadening of the resonance (increase of $\kappa$) accompanying the frequency fluctuations, which would be expected for a quasiparticle related origin (see supplementary materials for details). 

The strong dependence of the RTS amplitude on the sheet resistance suggests that the origin of the RTS lies in the granular structure of the film, i.e., the interplay between the Josephson coupling and the Coulomb repulsion. While more exotic TLS and quasiparticle processes have been found in highly disordered samples approaching the SIT \cite{Grunhaupt2018, deGraaf2020, Roy2020, Barone2018, Barone2020}, they are also subjected to the concerns brought forward above. 

Another mechanism that becomes relevant in the studied regime are collective modes of the superconducting condensate, i.e., fluctuations of the order parameter $\varPsi$ \cite{Raychaudhuri2021}. In particular, it has been shown theoretically that for a strongly disordered superconductor, phase modes acquire a dipole moment and appear below the gap, where they can have experimentally relevant lifetimes \cite{Cea2014}. Evidence of such modes in granular aluminum has been found in THz spectroscopy \cite{Pracht2017, Bertrand2019} and STM measurements \cite{Yang2020}. Calculations based on the bosonic model of the SIT showed that some modes even extend down to zero frequency where they can be thermally exited. This leads to fluctuations also in higher energy modes due to mode-mode interaction \cite{Feigelman2018}. However, theoretical frameworks describing the behavior of collective modes more precisely are still under development \cite{Khvalyuk2021}.

In conclusion, we have studied the low frequency excess noise in highly disordered granular aluminum resonators.
Our findings demonstrate that microwave resonator circuits are a valuable tool for the investigation of the SIT in disordered superconductors. The spectral analysis of the data suggests fluctuations of an RTS nature. While the amplitude $I$ of the RTS shows no dependence on the measurement power or the sample temperature, the RTS lifetime $\tau_0$ strongly decreases above a temperature of 200\,mK. Our data shows a correlation of both RTS amplitude $I$ and lifetime $\tau_0$ with the sheet resistance of the film. The measured absolute values and dependencies suggest that neither TLS nor quasiparticles cause the RTS. Instead, processes related to the reduced inter-grain coupling near the SIT seem to be a more likely explanation for the observed behavior. 

In comparison with other superconducting resonators, the measured frequency fluctuations are evidently linked with the nature of the granular material. Until a better understanding (and mitigation) of the physical origin of the RTS fluctuations and $1/f$ excess noise is available, highly resistive granular aluminum films close to the SIT are likely to introduce additional noise in superconducting circuits and detectors. 

The authors thank T. Wolz, M. Spiecker, J. Lisenfeld, M. Feigel'man and J. Cole for helpful discussions and L. Radtke for technical support. Samples were fabricated in the KIT Nanostructure Service Laboratory (NSL). This work was supported by the German Federal Ministry of Education and Research (GeQCoS and QSolid). The authors acknowledge partial support from the Landesgraduiertenf\"orderung of the state Baden-W\"urttemberg (M.W.) and the Helmholtz International Research School for Teratronics (J.N.V.).

\bibliographystyle{apsrev4-2}
\bibliography{bib}

\onecolumngrid
\newpage

\begin{center}
	\textbf{\large Supplementary Materials}
\end{center}

\setcounter{figure}{0}
\renewcommand{\figurename}{FIG.}
\renewcommand{\thefigure}{S\arabic{figure}}

\renewcommand{\tablename}{TAB.}
\renewcommand{\thetable}{S\arabic{table}}

\renewcommand{\theequation}{S\arabic{equation}}

\section{Measurement Setup}
\begin{figure}[h]
	\includegraphics{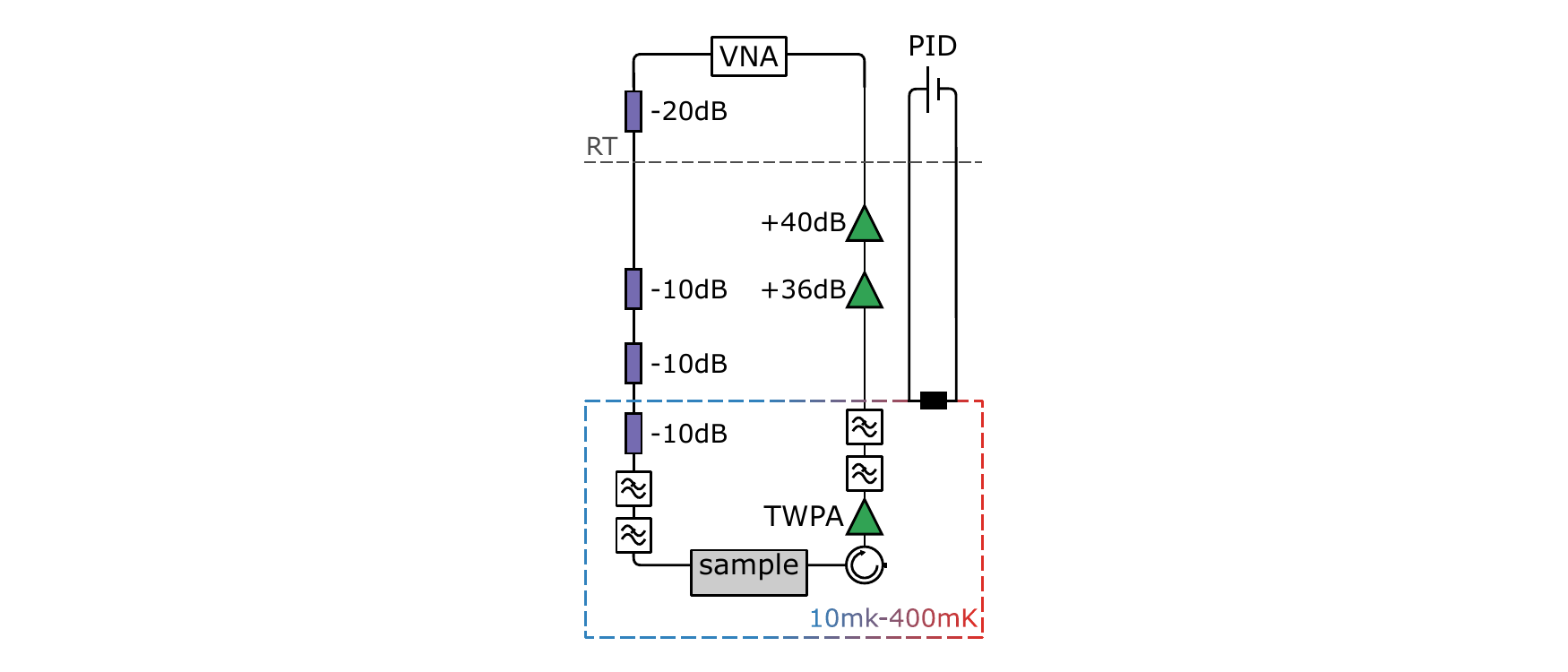}
	\caption{\label{FigS3} Schematic of the experimental setup and wiring including an attenuation and amplification chain with the corresponding components mounted to the various temperature stages of a dilution refrigerator. The temperature of the base plate, where the sample is installed, is regulated by a PID controller.}
\end{figure}

A schematic of the experimental setup is provided in Fig~\ref{FigS3}. All measurements of the complex transmission $S_{21}$ are performed with a commercial vector network analyzer (VNA). Coming from the VNA, the readout signal is attenuated  multiple times before reaching the sample mounted at the base plate of a dry dilution refrigerator. The temperature of the base plate can be controlled via electrical resistance heating. After leaving the sample, the signal passes a superconducting travelling wave parametric amplifier (TWPA) and two high-electron-mobility transistors (HEMT), which allows for measurements down to the few photon power limit, see Fig~3 in the main manuscript. Appropriate low pass filters protect the sample from infrared radiation.

\section{Frequency tracking and quasiparticle trajectory}

\begin{figure}[tbh]
	\includegraphics{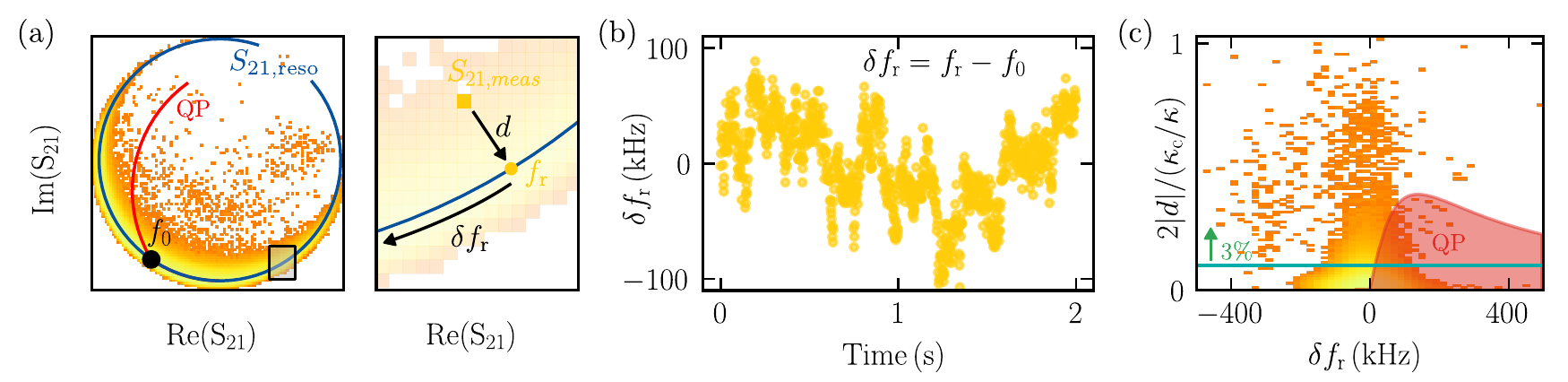}
	\caption{\label{FigS1} Measurement scheme for resonance frequency fluctuations. (a) Each raw data point $S_\textrm{21, meas}(f_0)$ is mapped on a prerecord resonance circle $S_\textrm{21, reso}(f)$ with radius $\kappa_{\rm c}/2\kappa$ to extract the resonator frequency $f_\textrm{r}$ corresponding to the moment of the recording. (b) Extract fluctuations $\delta f_\textrm{r} = f_\textrm{r} - f_0$ as a function of the recording time. (c) Mapping distance  $d = |S_\textrm{21, reso}(f_0)-S_\textrm{21, meas}(f_0)|$ as a function of the resonator frequency shift. The extracted values deviate strongly from the distribution expected for frequency fluctuations due to quasiparticles (QP, red). Generally, large mapping distances ($2|d|/(\kappa_{\rm c}/\kappa) > 0.1$) are found for less than $3 \%$ of the data points.}
\end{figure}

In the following we describe the employed method to deduce the change in resonator frequency $\delta f_\textrm{r}$ from a single frequency measurement at a fixed frequency $f_0$. The approach is depicted in Fig.~\ref{FigS1}(a). 

In a first step, the complex response circle $S_\textrm{21, reso}(f)$ of the resonator under investigation is measured by taking a single trace at a suitable frequency span $f_0 \pm \epsilon$ around the average resonance frequency $f_\textrm{0}$ with a VNA. This is a rather slow measurement process, but by fitting a circle to the full dataset, one obtains a reference lookup table for the subsequent, fast measurement. There, the time traces for the data presented in the paper is recorded by measuring the complex transmission $S_\textrm{21, meas}$ of each resonator at a constant frequency ${\sim}f_\textrm{0}$. When recording at base temperature ($10 \, \rm mK$), a rate of 500/s was chosen. Because the characteristic time of the RTS becomes shorter at higher temperature, this rate was increased to 4000/s above $T \geq 200 \, \rm mK$.

For each time trace we take $\sim 10^6$ such readings, refer to Fig.~\ref{FigS1}(a) for a typical distribution plotted in the complex plane. Note that averaging over several resonator frequencies during the reading of a single data point places it inside the resonance circle. Each data point is then used to calculate the momentary resonator frequency $f_\textrm{r}$. Mathematically, $f_\textrm{r}$ is defined as the frequency that minimizes the expression

\begin{align}
	\min_{f_\mathrm{r} \in [f_\textrm{0}-\epsilon,f_\textrm{0}+\epsilon]} |S_\textrm{21, reso}(\mathrm{r})-S_\textrm{21, meas}(f_0)|.
	\label{EqS1}
\end{align}

In practices, the algorithm solving Eq. \ref{EqS1} simply projects each data point onto the resonance circle, choosing the minimum projection distance.  With the knowledge of reference circle, the frequency corresponding to that point on the circle ($f_\textrm{r}$) is then known. Finally, the frequency shift is given as $\delta f_\textrm{r} = f_\textrm{r} - f_0$. An exemplary time trace obtained this way is partially shown in  Fig.~\ref{FigS1}(b). \\

The described method has several advantages. Apart from the simplicity of the experimental setup, it allows to accurately measure frequency shifts on the order of the resonator linewdith. In comparison, the linear approximation of the phase method brakes down in this regime and yields ambiguous results, see e.g. \cite{Wolz2021}. An analysis of the complex signal also allows for an identification of dissipative processes. By checking for deviations from the resonance circle $S_\textrm{21, reso}(f)$, it is possible to uncover quasiparticle (QP) related processes that can produce a random telegraph signal (RTS), as suggested by \cite{deVisser2011,deVisser2012}. 
There, each frequency shift is accompanied by a change in the resonator linewidth, which can be understood from the Mattis-Bardeen theory \cite{Mattis1958}

\begin{align}
	\kappa = \kappa_0 + 4 \pi \delta f_\mathrm{r} \frac{\mathrm{Re}(\iota)}{\mathrm{Im}(\iota)},
	\label{EqS2}
\end{align}  

where $\iota \propto \delta \sigma/\delta n_\textrm{qp} $ is the change in the complex conductivity $\sigma$ due to pair breaking into quasiparticles with density $n_\textrm{qp}$ \cite{Gao2008a}. The QP trajectory following Eq. \ref{EqS2} is plotted in Fig.~\ref{FigS1}(a). It is apparent that the measured set of data points does not follow this trajectory. Another way to see this is presented in Fig.~\ref{FigS1}(c), where the projection distance $d = |S_\textrm{21, reso}(f_0)-S_\textrm{21, meas}(f_0)|$ is plotted over the corresponding frequency shifts. A comparison between the measured (yellow) and QP distribution (red) indicates only a limited agreement for values around $\delta f_\textrm{r} \geq 120 \, \rm kHz$. 
As indicated by the horizontal line, where $|d|$ equals 10\% of the maximum mapping distance $\kappa_{\rm c}/2\kappa$ (= radius of resonance circle), potential QP events, together with all other measurement points in the center of the circle, only make up a negligible percentage ($\sim 3\%$) of the data. This suggests that QP may not be responsible for the broad Lorentzian RTS signature or excess 1/f noise spectrum.

\section{Allan analysis}  
\begin{figure}[tbh]
	\includegraphics{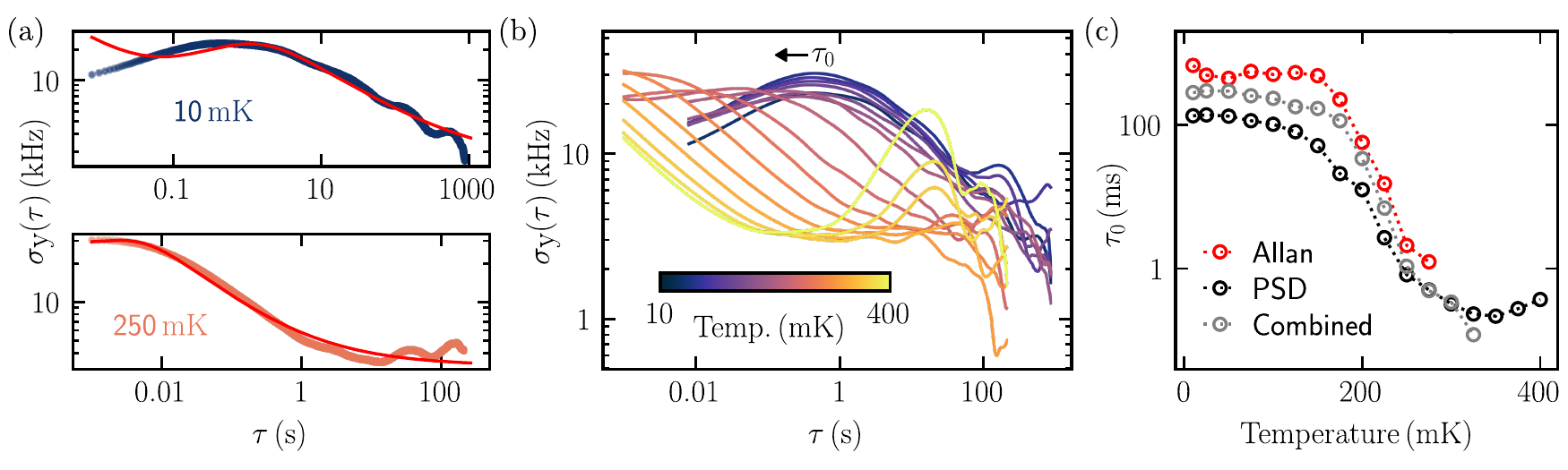}
	\caption{\label{FigS2} Noise analysis using the Allan deviation. (a) Allan deviation $\sigma_\textrm{y} (\tau)$ of the resonator fluctuations at 10\,mK and 250\,mK. The red line is a fit to Eq. \ref{EqS3}. (b) Continuous spectrum of Allan deviations. For increasing temperatures, the peak corresponding to the RTS lifetime $\tau_0$ shifts to the right. (c) Comparison between fits to the Allan deviation and the power spectral density (PSD). Above $275 \, \rm mK$, a second peak appears and the model no longer sufficiently describes the data. }
\end{figure}

As described in the main manuscript, the measurement data is evaluated via the power spectral density (PSD). An alternative, equally powerful tool is the (overlapping) Allan deviation $\sigma_\textrm{y} (\tau)$. If a time series is divided into adjacent segments $f_\textrm{k}$ of duration $\tau$, its Allan deviation is defined as $\sigma_\textrm{y} (\tau)= (\langle (\overline{f}_\textrm{k+1} - \overline{f}_\textrm{k})^2 \rangle /2)^{1/2}$. It is directly related to the power spectral density via integration, which allows us to write the Allan deviation equivalent of the noise spectrum model (Eq.~(1) in the main manuscript) as \cite{VanVliet1982, Niepce2021}

\begin{align}
	\sigma_\textrm{y} (\tau) = \frac{I\tau_0}{\tau} \left(4 \textrm{e}^{-\tau/\tau_0}-\textrm{e}^{-2\tau/\tau_0}+2\frac{\tau}{\tau_0}-3\right)^{1/2}+ \sqrt{2 h_{-1}\log(2)}+\sqrt{\frac{h_0}{2\tau}},
	\label{EqS3}
\end{align} 

where the first, second and third term describes a RTS, 1/f and white noise respectively. As shown in Fig.~\ref{FigS2}(a), this seemingly more complicated expression can properly fit the data at various temperatures. 

The advantage of the Allan deviation is the clear separation of the RTS peak from the noise background. This is apparent in Fig.~\ref{FigS2}(b), where $\sigma_\textrm{y} (\tau)$ is plotted for data measured on resonator C1 at different temperatures. There, the temperature dependent shift of $\tau_0$ can be easily observed in the noise spectrum. Fig.~\ref{FigS2}(c) compares the values extracted for $\tau_0$ when fitting either the Allan deviation, the PSD or both combined. The data shows a good qualitative agreement with the PSD data presented in the main manuscript.

At low temperatures the fits to Eq. \ref{EqS3} does not converge properly due to a secondary peak appearing on the right side of $\sigma_\textrm{y} (\tau)$ (Fig.~\ref{FigS2}(b)). The amplitude of these peaks increases with temperature while the corresponding lifetime decreases. At temperatures above 300\,mK it is reasonable to assume that thermally activated quasiparticle play a more prominent role. It is therefore likely that this is the origin of the secondary peaks.

\end{document}